\documentclass[12pt,english,aip, reprint]{revtex4-1}
\usepackage[T1]{fontenc}
\usepackage[latin9]{inputenc}
\usepackage{refstyle}
\usepackage{amsbsy}
\usepackage{amstext}

\makeatletter

\AtBeginDocument{\providecommand\secref[1]{\ref{sec:#1}}}
\AtBeginDocument{\providecommand\eqref[1]{\ref{eq:#1}}}
\AtBeginDocument{\providecommand\subsecref[1]{\ref{subsec:#1}}}
\AtBeginDocument{\providecommand\tabref[1]{\ref{tab:#1}}}
\providecommand{\tabularnewline}{\\}
\RS@ifundefined{subsecref}
  {\newref{subsec}{name = \RSsectxt}}
  {}
\RS@ifundefined{thmref}
  {\def\RSthmtxt{theorem~}\newref{thm}{name = \RSthmtxt}}
  {}
\RS@ifundefined{lemref}
  {\def\RSlemtxt{lemma~}\newref{lem}{name = \RSlemtxt}}
  {}

\usepackage{braket}
\usepackage[version=3]{mhchem}
\usepackage{url}
\usepackage{notes2bib}

\makeatother

\usepackage{babel}
\begin{document}
\newcommand*\citeref[1]{ref. \citenum{#1}}
\newcommand*\citerefs[1]{refs. \citenum{#1}} 

\newcommand*\Erkale{{\sc Erkale}}
\newcommand*\Qchem{{\sc Q-Chem}}
\newcommand*\poplebasis{\mbox{6-31(1+,3+)G*}}
\newcommand*\watercluster{\ce{(H2O)24-}}
\newcommand*\isomer{$4^{14}6^{4}$}
\title{Communication: Curing basis set overcompleteness with pivoted Cholesky
decompositions}
\author{Susi Lehtola}
\affiliation{Department of Chemistry, University of Helsinki, P.O. Box 55 (A. I. Virtasen
aukio 1), FI-00014 University of Helsinki, Finland}
\email{susi.lehtola@alumni.helsinki.fi}

\selectlanguage{english}%
\begin{abstract}
The description of weakly bound electronic states is especially difficult
with atomic orbital basis sets. The diffuse atomic basis functions
that are necessary to describe the extended electronic state generate
significant linear dependencies in the molecular basis set, which
may make the electronic structure calculations ill-convergent. We
propose a method where the over-complete molecular basis set is pruned
by a pivoted Cholesky decomposition of the overlap matrix, yielding
an optimal low-rank approximation that is numerically stable; the
pivot indices determining a reduced basis set that is complete enough
to describe all the basis functions in the original over-complete
basis. The method can be implemented either by a simple modification
to the usual canonical orthogonalization procedure, which hides the
excess functions and yields fewer efficiency benefits, or by generating
custom basis sets for all the atoms in the system, yielding significant
cost reductions in electronic structure calculations. The pruned basis
sets from the latter choice allow accurate calculations to be performed
at a lower cost even at the self-consistent field level, as illustrated
on a solvated \watercluster{} anion. Our results indicate that the
Cholesky procedure allows one to perform calculations with accuracies
close to standard augmented basis sets with cost savings which increase
with the size of the basis set, ranging from 9\% fewer functions in
single-$\zeta$ basis sets to 28\% fewer functions in triple-$\zeta$
basis sets.
\end{abstract}
\maketitle
\global\long\def\ERI#1#2{(#1|#2)}%
\global\long\def\bra#1{\Bra{#1}}%
\global\long\def\ket#1{\Ket{#1}}%
\global\long\def\braket#1{\Braket{#1}}%

\section{Introduction\label{sec:Introduction}}

Atomic orbital basis sets are a favorite choice in quantum chemistry,
as they afford a combination of speed and accuracy ranging from quick
and qualitative computations to ones nearing chemical accuracy at
a higher cost. As I have recently reviewed in \citeref{Lehtola2019c},
several kinds of atomic basis sets are commonly used: Gaussian-type
orbitals, Slater-type orbitals, as well as numerical atomic orbitals,
but the main advantage of all three is the same: a description based
on atomic orbitals tends to result in a systematic error to \emph{e.g.}
different spin states and at different geometries, which in many cases
results in fortuitous error cancellation for the reproduction of relative
energies, for example. 

One of the major stumbling blocks of atomic orbital based electronic
structure calculations is the description of the diffuse parts of
the wave function. An accurate portrayal requires atomic basis functions
with a large spatial extent that generate linear dependencies among
basis functions centered on different nuclei. This is especially an
issue in studies of loosely bound electrons,\citep{Herbert2015} which
frequently occur in dipole-\citep{Jordan2003} and quadrupole-bound\citep{Sommerfeld2014,Liu2019}
anions. Due to the weak binding, the outermost electron is delocalized
over a large region of space, which is hard to describe with atomic
basis functions as several shells of diffuse functions may be necessary
per atom, which in turn make the basis set overcomplete. Rydberg excited
states are likewise often extended, requiring far-reaching basis functions
for a proper portrayal,\citep{Jungen1981,Kaufmann1989} which may
again result in problems with linear dependencies in the molecular
basis set.

Benign linear dependencies can be removed with the canonical orthonormalization
procedure,\citep{Lowdin1956,Lowdin1970} in which eigenvectors of
the overlap matrix $S_{\mu\nu}=\braket{\mu|\nu}$ with small eigenvalues
are removed from the variational space. But, if the basis set is too
over-complete, the eigendecomposition of $\boldsymbol{S}$ is no longer
numerically stable, which prevents reliable calculations from taking
place.

Despite its pronounced importance, the over-completeness problem has
not been intensively studied in the literature. Indeed, a solution
has only been suggested for the special case of bond functions in
diatomic molecules.\citep{VanMourik1997} Instead, when faced with
problems with over-completeness, the established practice is to be
more judicious in the choice of the atoms where the diffuse functions
are placed, or to just settle for a smaller basis set; one widely
used example of the latter method are the calendar basis sets that
are only minimally augmented.\citep{Papajak2011} However, the former
method may be painstaking, and the latter method does not offer proof
of convergence to the complete basis set limit. 

In the present work, we propose an automated method which removes
the significant linear dependencies from a given over-complete molecular
basis set that is controlled by a single parameter. The method is
described in \secref{Method}, and results of the application thereof
to \watercluster{} are presented in \secref{Results}. The article
concludes with a summary and discussion in \secref{Summary-and-Discussion}. 

\section{Method\label{sec:Method}}

Our method is based on a pivoted Cholesky decomposition of the overlap
matrix as
\begin{equation}
S_{\mu\nu}\approx\tilde{S}_{\mu\nu}=\sum_{P=1}^{N}L_{\mu P}L_{\nu P}.\label{eq:cholesky}
\end{equation}
Although Cholesky decompositions have been previously used for basis
set orthogonalization in the context of linear-scaling calculations,\citep{Millam1997,Challacombe1999,Shao2003}
only complete decompositions appear to have been employed; the stability
and potential sparsity benefits afforded by an pivoted incomplete
algorithm do not appear to have been fully appreciated before now. 

The pivoted Cholesky decomposition, which has become commonly used
in various aspects of electronic structure theory,\citep{Aquilante2011}
has been shown to produce optimal low-rank representations of positive
semi-definite matrices,\citep{Harbrecht2012} making it suitable for
the present purposes. Starting from a large augmented basis set, the
pivoting picks those atomic basis functions that yield the most variational
freedom in the molecule, as judged by the basis function overlap.
The decomposition is continued until a predefined tolerance $\tau$
has been achieved, yielding at the end $\text{Tr }(\boldsymbol{S}-\tilde{\boldsymbol{S}})\leq\tau$.\citep{Harbrecht2012}
At the limit $\tau\to0$, all basis functions of the original over-complete
basis set can be represented in the truncated basis set, and so the
error of the approximation can be systematically removed.

Since the basis functions are typically normalized, $\braket{\mu|\mu}=1$,
at the beginning the Cholesky procedure does not know which basis
functions are truly important and which ones are not. This problem
does not occur in the Cholesky decomposition of the electron repulsion
integrals,\citep{Beebe1977} as tight functions have large self-repulsion
and thereby end up treated first by the algorithm. Because tight basis
functions cause the least issues with linear dependencies, we decided
to pick the initial pivot based on increasing length scales of the
basis functions, as determined by the expectation value of $\braket{r^{2}}$
around the center of the basis function.

\subsection{A simple approach\label{subsec:Simple-approach}}

A simple way to implement the proposed basis set truncation procedure
is to modify an existing canonical orthogonalization procedure, wherein
the overlap matrix is diagonalized
\begin{equation}
\boldsymbol{S}=\boldsymbol{\Sigma}\boldsymbol{\Lambda}\boldsymbol{\Sigma}^{\text{T}}\label{eq:Sdiag}
\end{equation}
and variational degrees of freedom $\boldsymbol{X}$ are obtained
as
\begin{equation}
\boldsymbol{X}=\boldsymbol{\Sigma}'\boldsymbol{\Lambda}'^{-1/2}\label{eq:varspace}
\end{equation}
where only those eigenvectors $\boldsymbol{\Sigma}_{i}$ are included
whose eigenvalues $\lambda_{i}$ are greater than the predefined threshold
$\epsilon$, $\lambda_{i}\ge\epsilon$.

The Cholesky truncation can be implemented by performing the pivoted
Cholesky decomposition in \eqref{cholesky} to a predefined threshold
$\tau$, yielding pivot indices $\boldsymbol{p}$. The set of molecular
basis functions corresponding to $\boldsymbol{p}$ exhibits fewer
linear dependencies than the original basis set. Now, the canonical
procedure, \eqref{Sdiag, varspace}, is performed for the submatrix
$\tilde{S}_{ij}=S_{p_{i}p_{j}}$, yielding a set of orthonormal vectors
$\tilde{X}_{ic}$. The corresponding vectors in the full space are
obtained as $X_{p_{i}c}=\tilde{X}_{ic}$, where the rows of $\boldsymbol{X}$
that do not appear in the pivot are set to zero.

Although this approach is simple to implement, it does not yield cost
savings in self-consistent field (SCF) calculations unless density-based
screening is employed in the integrals engine, since the procedure
just hides the excess functions in the basis set. The number of molecular
orbitals is, however, reduced, which is of main importance for post-Hartree--Fock
calculations which may exhibit steep scaling in the number of virtual
orbitals: for instance, the coupled-cluster\citep{Cizek1966} singles
and doubles (CCSD) model\citep{Purvis1982} scales as $O(v^{4})$,
where $v$ is the number of virtual orbitals.

\subsection{An efficient approach\label{subsec:An-efficient-approach}}

A more efficient solution can be fashioned in lines of the work of
Koch and coworkers on repulsion integral algoritms\citep{Koch2003}
as well as of Aquilante \emph{et al.} on automatic generation of auxiliary
basis sets based on Cholesky decompositions.\citep{Aquilante2007a,Aquilante2009}
As most electronic structure programs manipulate basis functions one
angular momentum shell at a time, the Cholesky decomposition in \eqref{cholesky}
can be modified so that all functions on the shell corresponding to
the pivot index are added simultaneously. As now shells are either
fully included in the basis or deleted altogether, one obtains a custom
basis set for each atom in the system, exhibiting optimal performance
characteristics for electronic structure calculations.

Because the Cholesky procedure can significantly modify the basis
sets on individual atoms, the basis set pruning procedure of this
scheme is not compatible with initial guesses that assume balanced
basis sets on each individual atom, such the superposition of atomic
densities guess.\citep{Almlof1982,VanLenthe2006} Instead, projective
approaches such as a minimal-basis guess wave function should be used.
The procedure can, however, be combined with the superposition of
atomic potentials initial guess\citep{Lehtola2019} which considers
the system as a whole.

\section{Results\label{sec:Results}}

The pivoted Cholesky decomposition method described in \subsecref{An-efficient-approach}
for the generation of pruned molecular basis sets has been implemented
in the \Erkale{} program.\citep{Lehtola2012,erkale} We demonstrate
the method using the $4^{14}6^{4}$ isomer of \watercluster{} from
\citeref{Herbert2005}. The BHLYP density functional is employed for
the demonstration, as it has been found to closely reproduce coupled-cluster
reference values for the system.\citep{Herbert2005} The functional
consists of half of Hartree--Fock and half of the local density exchange
functional,\citep{Bloch1929,Dirac1930} combined with the Lee--Yang--Parr
correlation functional.\citep{Lee1988} Single-point calculations
were performed in a development version of \Qchem{}\citep{Shao2015}
5.2 with a (100,590) quadrature grid, a $10^{-6}$ basis set linear
dependence threshold with a $10^{-16}$ threshold for the formation
of the overlap matrix, and a $10^{-14}$ screening threshold for two-electron
integrals. The wave functions were converged to an orbital gradient
threshold of $10^{-7}$, starting from wave functions converged to
$10^{-4}$ in the pc-0 basis.

The results of calculations in the special \poplebasis{} basis set
used in \citeref{Herbert2005}, obtained from the \mbox{6-31++G*}
basis set by adding two diffuse $s$ functions on hydrogen with a
progression factor $1/3$, are shown in \tabref{pople}. The Cholesky
truncation procedure allows one to reach the electron affinity predicted
by the full \poplebasis{} molecular basis set within 0.01 eV with
108 fewer basis functions, indicating a 15.5\% savings in the size
of the basis set required. Interestingly, the number of basis functions
that can be removed in this case, 108, is far more the number of linearly
dependent functions detected by the canonical orthogonalization procedure,
10. Furthermore, the truncated basis set contains no linear dependencies,
indicating that it is numerically better conditioned. No significant
differences can be seen in the number of SCF iterations required to
converge the wave function in the calculations with various truncations
of the \poplebasis{} basis set.

We have also studied the performance of the procedure with the single-
to triple-$\zeta$ augmented polarization consistent (pc) aug-pc-$n$
basis sets\citep{Jensen2001,Jensen2002b} as well as their doubly
(daug) and triply (taug) augmented versions obtained via geometric
extrapolation with \Erkale{}, the results of which are shown in \tabref{augpc}.
From these results it appears that doubly augmented pc basis sets
are sufficient to describe the weak binding of the solvated electron
in this system, and that a $\tau=10^{-6}$ decomposition threshold
affords a 0.01 eV accuracy for the detachment energy. While the pruning
only results in savings of 55 functions (-9\%) in the single-$\zeta$
basis, the savings increase in bigger basis sets to 292 functions
(-21\%) in the double-$\zeta$ and to 851 functions (-28\%) in the
triple-$\zeta$ calculations. Because larger atomic basis sets induce
more linear dependencies, it is likely that the savings in quadruple-$\zeta$
and higher basis sets would be even larger.

Significant differences in the number of SCF iterations can be seen
in the calculations with the larger basis sets. SCF appears to converge
smoothly in the low-threshold Cholesky truncated basis sets, whereas
more iterations are required in the original, overcomplete basis sets.
An extreme case is the triply augmented triple-$\zeta$ taug-pc-2
basis set, where the Cholesky truncations demonstrate a converged
electron affinity but the calculation in the original, pathologically
overcomplete basis set fails to converge for the anion.

\begin{table*}
\begin{centering}
\begin{tabular}{ccccr@{\extracolsep{0pt}.}lr@{\extracolsep{0pt}.}lr@{\extracolsep{0pt}.}lr@{\extracolsep{0pt}.}lc}
$\tau$ & $N_{\text{bf}}$ & $\min_{i}\lambda_{i}$ & $N_{\text{lin}}$ & \multicolumn{2}{c}{$E_{\text{neutral}}$ ($E_{h}$)} & \multicolumn{2}{c}{$N_{\text{neutral}}^{\text{SCF}}$} & \multicolumn{2}{c}{$E_{\text{anion}}$ ($E_{h}$)} & \multicolumn{2}{c}{$N_{\text{anion}}^{\text{SCF}}$} & $\Delta E$ (eV)\tabularnewline
\hline 
\hline 
$10^{-3}$ & 529 & $5.2\times10^{-3}$ & 529 & -1823&270626 & \multicolumn{2}{c}{10} & -1823&287328 & \multicolumn{2}{c}{12} & 0.45\tabularnewline
$10^{-4}$ & 588 & $9.6\times10^{-4}$ & 588 & -1823&315169 & \multicolumn{2}{c}{10} & -1823&330750 & \multicolumn{2}{c}{13} & 0.42\tabularnewline
$10^{-5}$ & 628 & $2.1\times10^{-4}$ & 628 & -1823&322834 & \multicolumn{2}{c}{10} & -1823&338134 & \multicolumn{2}{c}{13} & 0.42\tabularnewline
$10^{-6}$ & 657 & $2.4\times10^{-5}$ & 657 & -1823&327639 & \multicolumn{2}{c}{10} & -1823&343031 & \multicolumn{2}{c}{12} & 0.42\tabularnewline
$10^{-7}$ & 672 & $3.6\times10^{-6}$ & 672 & -1823&328289 & \multicolumn{2}{c}{10} & -1823&343814 & \multicolumn{2}{c}{12} & 0.42\tabularnewline
$10^{-8}$ & 685 & $7.4\times10^{-7}$ & 684 & -1823&328435 & \multicolumn{2}{c}{10} & -1823&343960 & \multicolumn{2}{c}{12} & 0.42\tabularnewline
w/o & 696 & $5.4\times10^{-8}$ & 686 & -1823&328446 & \multicolumn{2}{c}{10} & -1823&343971 & \multicolumn{2}{c}{12} & 0.42\tabularnewline
\end{tabular}
\par\end{centering}
\caption{Vertical electron detachment energies of the \isomer{} isomer of
\watercluster{} at the geometry from \citeref{Herbert2005} with
the BHLYP functional and the \poplebasis{} basis set. Column legend:
Cholesky decomposition threshold $\tau$, total number of basis functions
$N_{\text{bf}}$, smallest eigenvalue of the overlap matrix $\min_{i}\lambda_{i}$,
number of linearly independent basis functions $N_{\text{lin}}$,
total energy and number of SCF iterations for neutral cluster ($E_{\text{neutral}}$
and $N_{\text{neutral}}^{\text{SCF}}$) and cluster anion ($E_{\text{anion}}$
and $N_{\text{anion}}^{\text{SCF}}$), and the resulting electron
affinity $\Delta E$. The last row shows the results without (w/o)
the Cholesky truncation procedure.\label{tab:pople}}
\end{table*}

\begin{table*}
\begin{centering}
\begin{tabular}{c|ccccc|ccccc|ccccc}
 & \multicolumn{5}{c|}{aug-pc-0} & \multicolumn{5}{c|}{daug-pc-0} & \multicolumn{5}{c}{taug-pc-0}\tabularnewline
$\tau$ & $N_{\text{bf}}$ & $\min_{i}\lambda_{i}$ & $N_{\text{neutral}}^{\text{SCF}}$ & $N_{\text{anion}}^{\text{SCF}}$ & $\Delta E$ & $N_{\text{bf}}$ & $\min_{i}\lambda_{i}$ & $N_{\text{neutral}}^{\text{SCF}}$ & $N_{\text{anion}}^{\text{SCF}}$ & $\Delta E$ & $N_{\text{bf}}$ & $\min_{i}\lambda_{i}$ & $N_{\text{neutral}}^{\text{SCF}}$ & $N_{\text{anion}}^{\text{SCF}}$ & $\Delta E$\tabularnewline
\hline 
\hline 
$10^{-3}$ & 390 & $9.2\times10^{-3}$ & 10 & 12 & 0.64 & 425 & $2.4\times10^{-3}$ & 10 & 12 & 0.65 & 442 & $1.9\times10^{-3}$ & 10 & 12 & 0.63\tabularnewline
$10^{-4}$ & 426 & $2.3\times10^{-3}$ & 10 & 11 & 0.60 & 472 & $7.5\times10^{-4}$ & 10 & 12 & 0.61 & 500 & $1.8\times10^{-4}$ & 10 & 12 & 0.61\tabularnewline
$10^{-5}$ & 447 & $7.1\times10^{-4}$ & 10 & 12 & 0.69 & 509 & $8.1\times10^{-5}$ & 10 & 12 & 0.68 & 546 & $2.0\times10^{-5}$ & 9 & 12 & 0.68\tabularnewline
$10^{-6}$ & 456 & $2.7\times10^{-4}$ & 10 & 12 & 0.69 & 545 & $3.3\times10^{-6}$ & 9 & 12 & 0.68 & 594 & $1.3\times10^{-6}$ & 11 & 14 & 0.68\tabularnewline
$10^{-7}$ & 456 & $2.7\times10^{-4}$ & 10 & 12 & 0.69 & 579 & $2.7\times10^{-6}$ & 9 & 12 & 0.67 & 646 & $2.0\times10^{-7}$ & 11 & 14 & 0.67\tabularnewline
$10^{-8}$ & 456 & $2.7\times10^{-4}$ & 10 & 12 & 0.69 & 589 & $8.5\times10^{-7}$ & 9 & 12 & 0.68 & 672 & $1.0\times10^{-7}$ & 11 & 15 & 0.67\tabularnewline
w/o & 456 & $2.7\times10^{-4}$ & 10 & 12 & 0.69 & 600 & $1.4\times10^{-7}$ & 9 & 12 & 0.68 & 744 & $2.0\times10^{-11}$ & 11 & 15 & 0.68\tabularnewline
\multicolumn{1}{c}{} &  &  &  &  & \multicolumn{1}{c}{} &  &  &  &  & \multicolumn{1}{c}{} &  &  &  &  & \tabularnewline
 & \multicolumn{5}{c|}{aug-pc-1} & \multicolumn{5}{c|}{daug-pc-1} & \multicolumn{5}{c}{taug-pc-1}\tabularnewline
$\tau$ & $N_{\text{bf}}$ & $\min_{i}\lambda_{i}$ & $N_{\text{neutral}}^{\text{SCF}}$ & $N_{\text{anion}}^{\text{SCF}}$ & $\Delta E$ & $N_{\text{bf}}$ & $\min_{i}\lambda_{i}$ & $N_{\text{neutral}}^{\text{SCF}}$ & $N_{\text{anion}}^{\text{SCF}}$ & $\Delta E$ & $N_{\text{bf}}$ & $\min_{i}\lambda_{i}$ & $N_{\text{neutral}}^{\text{SCF}}$ & $N_{\text{anion}}^{\text{SCF}}$ & $\Delta E$\tabularnewline
\hline 
\hline 
$10^{-3}$ & 743 & $5.4\times10^{-3}$ & 10 & 11 & 0.20 & 834 & $1.2\times10^{-3}$ & 10 & 12 & 0.29 & 870 & $7.1\times10^{-4}$ & 10 & 16 & 0.30\tabularnewline
$10^{-4}$ & 813 & $8.4\times10^{-4}$ & 10 & 12 & 0.27 & 917 & $1.6\times10^{-4}$ & 10 & 12 & 0.33 & 979 & $1.0\times10^{-5}$ & 10 & 16 & 0.33\tabularnewline
$10^{-5}$ & 882 & $1.9\times10^{-4}$ & 10 & 12 & 0.29 & 1005 & $5.4\times10^{-5}$ & 10 & 12 & 0.34 & 1082 & $8.8\times10^{-6}$ & 10 & 16 & 0.34\tabularnewline
$10^{-6}$ & 943 & $7.0\times10^{-5}$ & 10 & 12 & 0.33 & 1100 & $8.2\times10^{-6}$ & 10 & 12 & 0.36 & 1206 & $5.5\times10^{-7}$ & 10 & 15 & 0.36\tabularnewline
$10^{-7}$ & 957 & $1.4\times10^{-5}$ & 10 & 12 & 0.33 & 1167 & $8.4\times10^{-7}$ & 10 & 12 & 0.36 & 1287 & $2.0\times10^{-7}$ & 13 & 21 & 0.36\tabularnewline
$10^{-8}$ & 980 & $9.2\times10^{-6}$ & 10 & 12 & 0.34 & 1210 & $1.9\times10^{-7}$ & 10 & 12 & 0.36 & 1350 & $8.1\times10^{-9}$ & 13 & 23 & 0.36\tabularnewline
w/o & 984 & $6.5\times10^{-6}$ & 10 & 12 & 0.34 & 1392 & $1.7\times10^{-12}$ & 12 & 16 & 0.37 & 1800 & $-1.5\times10^{-15}$ & 14 & 23 & 0.37\tabularnewline
\multicolumn{1}{c}{} &  &  &  &  & \multicolumn{1}{c}{} &  &  &  &  & \multicolumn{1}{c}{} &  &  &  &  & \tabularnewline
 & \multicolumn{5}{c|}{aug-pc-2} & \multicolumn{5}{c|}{daug-pc-2} & \multicolumn{5}{c}{taug-pc-2}\tabularnewline
$\tau$ & $N_{\text{bf}}$ & $\min_{i}\lambda_{i}$ & $N_{\text{neutral}}^{\text{SCF}}$ & $N_{\text{anion}}^{\text{SCF}}$ & $\Delta E$ & $N_{\text{bf}}$ & $\min_{i}\lambda_{i}$ & $N_{\text{neutral}}^{\text{SCF}}$ & $N_{\text{anion}}^{\text{SCF}}$ & $\Delta E$ & $N_{\text{bf}}$ & $\min_{i}\lambda_{i}$ & $N_{\text{neutral}}^{\text{SCF}}$ & $N_{\text{anion}}^{\text{SCF}}$ & $\Delta E$\tabularnewline
\hline 
\hline 
$10^{-3}$ & 1550 & $4.2\times10^{-3}$ & 10 & 11 & 0.25 & 1669 & $1.2\times10^{-3}$ & 10 & 11 & 0.34 & 1746 & $5.5\times10^{-4}$ & 9 & 13 & 0.36\tabularnewline
$10^{-4}$ & 1672 & $7.7\times10^{-4}$ & 10 & 11 & 0.36 & 1828 & $1.9\times10^{-4}$ & 10 & 11 & 0.38 & 1924 & $5.0\times10^{-6}$ & 10 & 12 & 0.39\tabularnewline
$10^{-5}$ & 1817 & $1.2\times10^{-4}$ & 10 & 11 & 0.33 & 2003 & $1.3\times10^{-5}$ & 10 & 11 & 0.35 & 2122 & $1.8\times10^{-6}$ & 10 & 13 & 0.36\tabularnewline
$10^{-6}$ & 1966 & $2.6\times10^{-5}$ & 10 & 10 & 0.32 & 2173 & $3.5\times10^{-6}$ & 10 & 11 & 0.34 & 2333 & $2.8\times10^{-7}$ & 10 & 23 & 0.34\tabularnewline
$10^{-7}$ & 2069 & $6.9\times10^{-6}$ & 10 & 10 & 0.32 & 2339 & $5.7\times10^{-7}$ & 11 & 15 & 0.34 & 2510 & $1.6\times10^{-7}$ & 12 & 32 & 0.34\tabularnewline
$10^{-8}$ & 2109 & $1.1\times10^{-6}$ & 10 & 12 & 0.32 & 2468 & $8.3\times10^{-8}$ & 12 & 16 & 0.34 & 2674 & $8.4\times10^{-9}$ & 15 & 24 & 0.34\tabularnewline
w/o & 2208 & $9.1\times10^{-9}$ & 12 & 14 & 0.33 & 3024 & $2.2\times10^{-15}$ & 14 & 17 & 0.34 & 3840 & $-3.1\times10^{-15}$ & 14 & \multicolumn{2}{c}{no convergence}\tabularnewline
\end{tabular}
\par\end{centering}
\caption{Vertical electron detachment energies of the \isomer{} isomer of
\watercluster{} at the geometry from \citeref{Herbert2005} with
the BHLYP functional and various polarization consistent basis sets.
The notation is the same as in \tabref{pople}.\label{tab:augpc}}
\end{table*}

\section{Summary and Discussion\label{sec:Summary-and-Discussion}}

We have suggested pivoted Cholesky decompositions as a way to overcome
numerical difficulties with overcomplete basis sets by explicit removal
of linearly dependent functions. In the proposed approach, the pivot
indices from an incomplete Cholesky factorization of the overlap matrix
are used to pick a subset of atomic orbitals that form a sufficiently
complete basis for the molecule. The scheme can be implemented either
by i) a simple modification to existing basis set orthogonalization
procedures that amounts to hiding the rest of the basis functions
(scheme presented in \subsecref{Simple-approach}), or ii) by constructing
pruned basis sets at each geometry by complete removal of unnecessary
shells of basis functions from the calculation (scheme presented in
\subsecref{An-efficient-approach}).

We have demonstrated the suitability, stability and efficiency of
the latter approach with calculations on a weakly bound anion, \watercluster{}.
We have found that the vertical detachment energy is reproduced within
0.01 eV for this system with a $\tau=10^{-6}$ decomposition threshold,
requiring 9\% to 28\% fewer basis functions than the full original
basis sets. As the eliminated functions are diffuse ones that generally
do not screen well in integral computations, the large number of deleted
functions implies significant savings in computer time.

The Cholesky decomposition approach is generally applicable to electronic
structure calculations in atomic basis sets regardless of their form:
in addition to the Gaussian basis sets used in the present work, the
algorithm can also be used in combination with Slater-type and numerical
atomic orbital basis sets. The procedure could especially be combined
with the Gaussian cell model,\citep{Haines1974,Ralston1995,Wilson1995a,Wilson1996a}
which has been recently resuggested as the off-center Gaussian model.\citep{Melichercik2013,Melichercik2018}
In addition, the decomposition procedure could also be used with auxiliary
basis sets for resolution of identity approaches.\citep{Vahtras1993}

Although only fixed geometries have been considered in the present
work, the extension to geometry optimization and potential energy
surfaces is straightforward. For either of the approaches, \subsecref{Simple-approach}
or \subsecref{An-efficient-approach}, the number of molecular orbitals
may depend on the geometry. However, the same issue also exists for
the canonical orthogonalization approach, and should not cause problems
with reasonably small thresholds. Note that in either case, \ref{subsec:Simple-approach}
or \ref{subsec:An-efficient-approach}, the pruned basis set should
be redetermined at each geometry.

\section*{Acknowledgments}

I thank Roland Lindh for discussions on the atomic auxiliary function
Cholesky method. This work has been supported by the Academy of Finland
through project number 311149. Computational resources provided by
CSC -- It Center for Science Ltd (Espoo, Finland) and the Finnish
Grid and Cloud Infrastructure (persistent identifier urn:nbn:fi:research-infras-2016072533)
are gratefully acknowledged.

\bibliography{citations}

\end{document}